\documentclass[a4paper,12pt]{article}
\usepackage{inputenc}
\usepackage{graphicx}
\usepackage{amsmath}
\usepackage{amsfonts}
\usepackage{amssymb}
\usepackage{textcomp}
\usepackage{xcolor}
\usepackage{float}
\usepackage{appendix} 

\def \Re{\mathop{\mbox{\normalfont{\bf{Re}}}}\nolimits}
\def \M{\mathcal{M}}

\def \Br{{\bf{Br}}}
\def \TeV{\mathop{\mbox{\normalfont TeV}}\nolimits}
\def \GeV{\mathop{\mbox{\normalfont GeV}}\nolimits}
\def \pb{\mathop{\mbox{\normalfont pb}}\nolimits}
\def \ab{\mathop{\mbox{\normalfont ab}}\nolimits}

\newcommand{\be}{\begin{eqnarray}}
\newcommand{\ee}{\end{eqnarray}}
\newcommand{\nn}{\nonumber}
\newcommand{\bma}{\begin{matrix}}
\newcommand{\ema}{\end{matrix}}

\catcode`\@=11
\def\lsim{\mathrel{\mathpalette\@versim<}}
\def\gsim{\mathrel{\mathpalette\@versim>}}
\def\@versim#1#2{\vcenter{\offinterlineskip
\ialign{$\m@th#1\hfil##\hfil$\crcr#2\crcr\sim\crcr } }}
\catcode`\@=12

\catcode`@=12

\begin{document}
\thispagestyle{empty}
\begin{flushright}
FCFM-HEP-001\\
March 2013\
\end{flushright}
\begin{center}
{\Large \bf An extra $Z'$ gauge boson as a source of Higgs particles
\\} 
\vspace{1.5in}
{\bf J. Lorenzo D\'iaz-Cruz$^1$, Javier M. Hern\'andez-L\'opez$^1$ and Javier A. Orduz-Ducuara$^1$}
\vspace{0.2in}\\
{\sl $^1$ Facultad de Ciencias F\'isico-Matem\'aticas, \\
Benem\'erita Universidad Aut\'onoma de Puebla, Puebla, M\'exico }
\end{center}
\begin{abstract}
Models with extra gauge bosons often require an extended Higgs sector, 
which contains a rich spectrum of Higgs bosons with properties that deviate from the Standard Model (SM). 
Such Higgs bosons could be searched using standard mechanisms similar to the SM.
However, the existence of the new gauge bosons could provide new production mechanisms, 
which could probe the non-standard origin of the Higgs particle. In this paper we study 
how the $Z'$ from a model with extra $U(1)'$ could
be used as a source of Higgs bosons. We study  3-bodies decays of the $Z'$ into 
a Higgs boson and a top anti-top pair or WW pair, namely  $Z'\to t\bar{t} h, WW h.$ 
We find that it is possible to get ${\Br'}$s as high as $10^{-2}$ for these modes, which could 
be studied at future colliders. We also study  the production of the 
Higgs bosons in association with the $Z$ vector boson at a linear collider,
through the reaction $e^+ e^- \to Z, Z'\to Z+h$, including both the resonant and the
non-resonant effects. 

\end{abstract}

\section{Introduction}
After many years of preparation, the LHC has tested in very significant
ways the Higgs sector of the SM. In fact 
the LHC announcement of the discovery of a new boson, with properties
that resemble those of the SM Higgs boson can be considered a great triumph of particle physics. 
The new  boson has a mass $m_h \simeq 125 \GeV$ 
\cite{CMS-bsn:2012gu,ATLAS-bsn:2012gk}, 
which agrees quite well with the range preferred by the analysis of
electroweak precision tests. The LHC has also searched for new physics beyond the SM, 
and so far only has extended the bounds on the new physics scale.

Some of the simplest extensions of the SM are those that add new gauge bosons, which could 
arise from a variety of contexts, ranging from simple extra $U(1)$ gauge symmetries \cite{Pati:1974yy}, 
left-right models  \cite{Pati:1974yy,Mohapatra:1974hk,Mohapatra:1980yp,Mohapatra:1979ia}, 
GUTs \cite{Kane:1998hz}, and even string theory \cite{Cvetic:1995rj, Cvetic:1996mf, Cleaver:1998sm}. 
In particular, extension of the SM with an extra $U(1)'$ have been studied extensively, 
including the detection at LHC, as well as other phenomenological aspects \cite{Pati:1974yy,Mohapatra:1974hk,Mohapatra:1980yp,Mohapatra:1979ia,Kane:1998hz,Cvetic:1995rj,
Cvetic:1996mf,Cleaver:1998sm}.

The crucial test for this kind of models is precisely the presence of new $Z'$
gauge boson, and recent bounds from LHC indicate that it should be heavier than
about $1.5 \TeV$\cite{langacker-09,Diener:2009vq,Hsieh:2010zr}. Future LHC runs at 13-14 TeV will increase the $Z'$ mass bounds or luckily will find
evidence of its presence. Further studies will require a new linear collider,
which will also allow us to perform precision studies of the Higgs sector, 
in order to search for the nature of the Higgs	 mechanism at more fundamental level.

The Higgs boson from these models could be searched using mechanisms similar to the SM one.
However, the existence of such new gauge bosons could also 
provide new Higgs production mechanisms, which could probe its non-standard origin.
In this paper we are interested in studying how the $Z'$ from a model with extra $U(1)'$, 
could be used as a source of Higgs bosons.  We calculate the decay widths for the 3-body 
decays of the $Z'$ into a Higgs boson, namely for $Z'\to t\bar{t} h, WW h$.  
We find that it is possible to get ${\Br}$'s as high 
as $10^{-2}$ for these modes,  which could be studied at LHC. We also calculate the production of the 
Higgs bosons in association with the $Z$ vector boson at a linear collider, 
i.e. $e^+ e^- \to Z, Z'\to Z+h$, with $Z'$ either on-shell or off-shell. 

The organization of this paper goes as follows: section 2 contains a summary of the
$U(1)_{B-L}$ model with extra $Z'$. Then, section 3 includes the calculation of the
3-body decays of $Z'$ into a Higgs boson and a pair of SM particles, $Z'\to t\bar{t} h, WW h$. 
Section 4 contains our calculation of $e^+ e^- \to Z, Z'\to Z+h$, while Section 5
includes our conclusions. Some of the lenghtly formulae appearing in the calculation of
the 3-body decays are included in the appendices.

\section{A $U(1)_{B-L}$ model with extra $Z':$ parameters and bounds}

In this section we will describe the Lagrangian, including the gauge and 
fermionic sectors for ${SU(2)_L \times {U(1)}_Y \times {U(1)'}}$ model, 
we also write the Higgs potential and the Higgs coupling. The gauge sector 
is given by\cite{Ferroglia:2006mj, Rizzo:2006nw},
\be\label{eq:Lagr-gral-Uup-gauge-states}
{{\mathcal{L}}}_{g}&=& -\frac{1}{4}{B}_{\mu\nu}{B}^{\mu\nu}-\frac{1}{4}
{W}^a_{\mu\nu}{W^a}^{\mu\nu} 
-\frac{1}{4} {Z'}_{\mu\nu} {Z'}^{\mu\nu} 
\ee
We ignore the $SU(3)$ sector for the purposes of the paper. The field strength 
tensor are labeled as $W^a_{\mu\nu}, B_{\mu\nu}$ and $Z'_{\mu\nu}$ for ${SU(2)}_L, {U(1)_Y}$ and $U(1)',$ 
respectively. We need to transform  eq.\eqref{eq:Lagr-gral-Uup-gauge-states} in order to obtain the physical mass 
eigenstates \cite{Foot:1991kb,Babu:1997st}.
	
For the scalar sector we have,
\be\label{ec:lagran-gral}
{\mathcal{L}}_s&=&\left(D^\mu \Phi\right)^{\dagger}\left(D_\mu \Phi\right)+\left(D^\mu \chi \right)^{\dagger}\left(D_\mu \chi\right)-V(\Phi,\chi)\nn
\ee
where $\Phi$ denotes the SM scalar doublet and $\chi$ is a singlet under the SM. The covariant derivate for 
a pure $B-L$ model is \cite{Basso:2010pe,Basso-Moretti-Pruna-2010,Basso:2010jm},
\be
	D_\mu&=&\partial_\mu+i\left[g T^a
	W^a_\mu+g_1YB_\mu+g'Y'B'_\mu\right]\nn
\ee
The doublet and  singlet scalars are further written as,
\be
\Phi=
  \begin{pmatrix}
G^\pm\\
\frac{v + \phi^0 + iG_Z}{\sqrt{2}}\\
  \end{pmatrix},
\hspace{5mm}
\chi = \frac{v' + {\phi'}^0 + iz'}{\sqrt{2}}
\ee
In this way, $G^\pm, G_Z$ and $z'$ become the Goldstone bosons of $W^\pm, Z$ and $Z',$ respectively. 
The potential that does this is \cite{Basso:2010jm},
\be
V(\Phi,\chi)&=&-m^2\Phi^\dag \Phi-\mu^2|\chi|^2+\lambda_1(\Phi^\dag \Phi)^2+\lambda_2|\chi|^4+\lambda_3\Phi^\dag \Phi|\chi|^2\nn
\ee
After SSB we obtain the mass matrix for the neutral CP-even states, $\phi^0$ and ${\phi'}^0$. Then, 
we need to diagonalize this mass matrix to obtain the mass eigenstates, i.e. 
\be
  \begin{pmatrix}
h\\
H\\
  \end{pmatrix} = 
  \begin{pmatrix}
\cos\alpha&-\sin\alpha\\
\sin\alpha&\cos\alpha\\
  \end{pmatrix}
  \begin{pmatrix}
\phi^0\\
{\phi'}^0\\
  \end{pmatrix}
\ee  

Where $\alpha$ is the mixing angle for the neutral Higgs bosons. However in order to respect the recent constraints from LHC on the Higgs couplings\cite{CMS-bsn:2012gu, ATLAS-bsn:2012gk}
we shall fix the mixing angle to be $\cos\alpha \simeq 1,$ i.e. $h \simeq {\phi}^0.$

The couplings of a $Z'$ boson to the SM fermions are given by,
{\small{
\be
{{\mathcal{L}}}_{NC}&=& g^{}_{Zff} \sum_{f}\bar{f} \gamma^\mu \bigg({g}_{V}^{f} + {g}_{A}^{f}\gamma^5\bigg)f {Z}_\mu+g^{}_{Zff} \sum_{f}\bar{f} \gamma^\mu \bigg({g'}_{V}^{f} + {g'}_{A}^{f}\gamma^5\bigg)f {Z'}_\mu\label{eq:L-Zff}
\ee
}}
where $g_{Zff} = \frac{g}{\cos\theta_W},$ 
\be\label{eq:gV-gA-SM}
\begin{array}{ll}
g_V^f &= \frac{T_3^f}{2} \cos\theta' - Q^f \cos\theta' \sin^2\theta_W - \frac{g'}{2g_1} \sin\theta' \sin\theta_W,\\
g_A^f&=-\frac{T_3^f}{2} \cos\theta'\cos^2\theta_W
\end{array}
\ee
and
\be\label{eq:gV-gA-SM-ext}
\begin{array}{ll}
{g'}_V^f &=\frac{T_3^f}{2}\sin\theta' - Q^f\sin\theta' \sin^2\theta_W + \frac{g'}{2g_1} \cos\theta' \sin\theta_W,\\
{g'}_A^f &=-\frac{T_3^f}{2}\sin\theta'\cos^2\theta_W
\end{array}
\ee
As it is given in the SM $\frac{g_1}{g} = \frac{\sin\theta_W}{\cos\theta_W}.$ Similarly,  in the forthcoming calculations we will 
use the following definitions the couplings for gauge and neutral Higgs bosons which are shown  in table (\ref{ta:ver-bos-higgs-fermio}).

\begin{table}[H]
\centering
\caption{Definitions for vertices with gauge and Higgs boson.}
{\begin{tabular}{cc} \hline
$g_{ZWW}$&
$g_{Z'WW}$\\
\hline
$ig\cos\theta'\sin\theta_W$&
$ig\sin\theta'\sin\theta_W$\\
\hline
\end{tabular} \label{ta:ver-bos-higgs-fermio}}
\end{table}
and 
\be
g_{hZZ' }=\frac{1}{4}\frac{v}{M_W}\Big[f(\theta')\cos\alpha
-
g(\theta')\sin\alpha\Big]\nn
\ee
where $f(\theta')=\sin2\theta' \big(g^2 + g_1^2 - {g'}^2\big) + 2g' \cos2\theta' \sqrt{g^2 + g_1^2},~~ g(\theta') = 4r  g'^2 
\sin2\theta'$ and $r = \frac{v'}{v}.$ The $Z-Z'$ mixing angle is taken as
$\theta' =  10^{-3}$ ($\theta' =  10^{-4}$) which respects the currents bounds on this 
parameter\cite{beringer-pdg-2012}. 
According to recent LHC results we must take $m_h = 125 \GeV,$ 
for the SM-like Higgs boson, the vacuum expectation value is taken  as
$v'=2 \TeV.$ We took the value $\alpha = \frac{\pi}{9}$ for the Higgs mixing
parameter, which is in the agreement with \cite{CMS-bsn:2012gu, ATLAS-bsn:2012gk, Chakraborty:2013si}.

\section{Three body decays: $Z'\to t\bar{t}h, W^+W^-h$}
In the following we present the formulae for the 3-body decays of the $Z'$ that include a Higgs 
boson in the final state, namely for the modes $Z'\to t\bar{t} h$ and $Z'\to WW h$. 
Since the dominant two-body decay modes are needed in order to
calculate the corresponding branching ratios (${\Br}$), we shall also present the explicit
expressions for the corresponding decay widths.

\subsection{Two-body modes}

The dominant decays of $Z'$ includes the decay into a fermion pair ($Z' \to f\bar{f}$), 
gauge boson pair ($Z' \to W^+W^-$) and into a higgs and a Z boson ($Z' \to Zh$).
\begin{figure}[H]
\vspace*{-2cm}
\includegraphics[width=6.5cm,angle =-90]{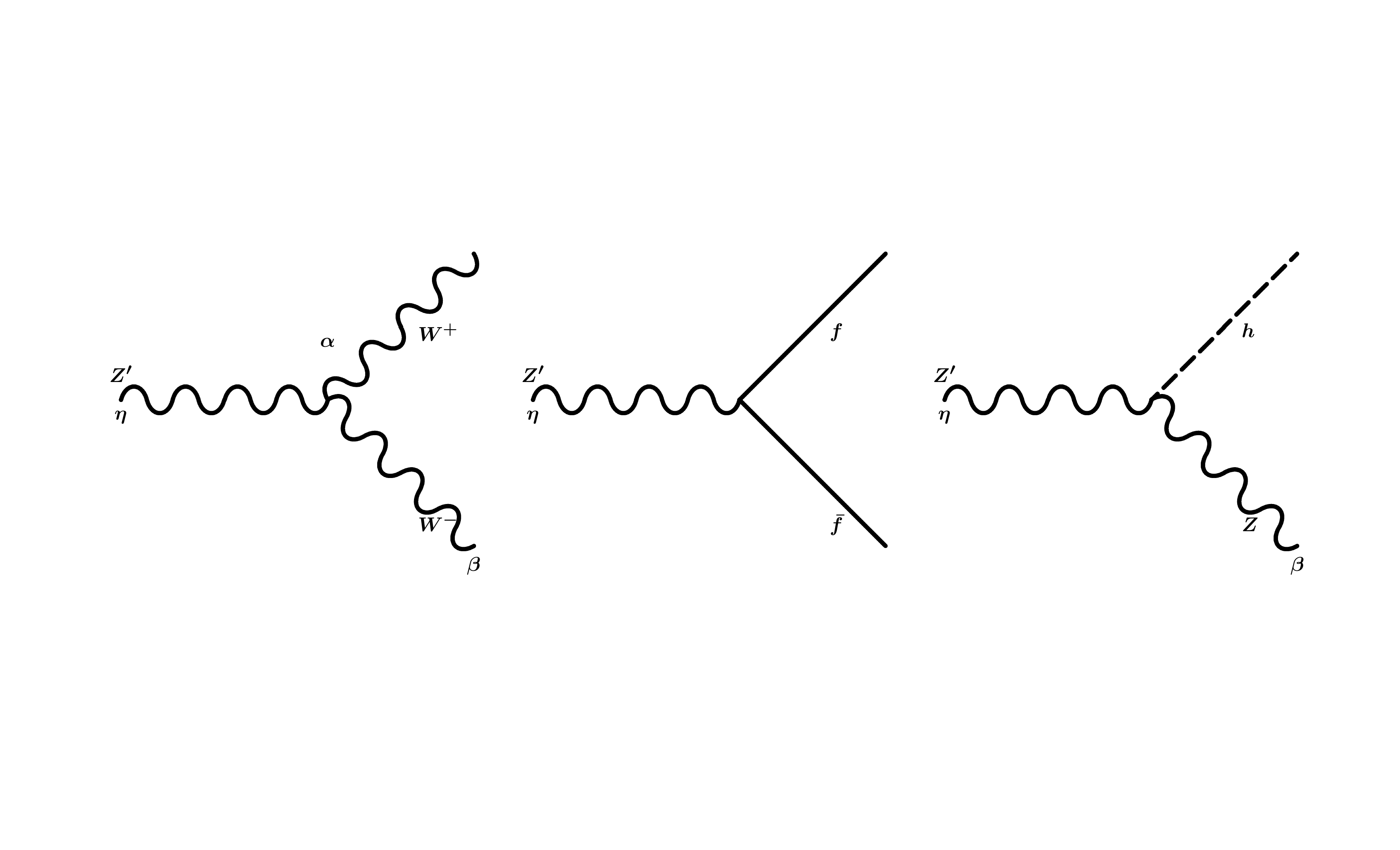}
\vspace*{-2cm}
\caption{Feynman diagrams for two bodies}
\end{figure}
The decay width for $Z' \to W^+W^-$ process is given by\cite{Leike:1998wr}:
\be\label{eq:Zp-WMWm}
\Gamma(Z'\to W^+W^-) =
\frac{g_{Z'WW}^2 M_{Z'}}{128 \pi r_{W}^2  }
(1-2 r_{W})^{1/2} (1-4 r_{W}) 
(r_{W} (32 r_{W}-5)+1)
\ee
where $g_{Z' W W}$ is include in table (\ref{ta:ver-bos-higgs-fermio}) and $r_W = \frac{M_W^2}{M_{Z'}^2}.$ 

The decays width for $Z' \to f\bar{f}$ is given by\cite{Ferroglia:2006mj, Leike:1998wr}:
\be\label{eq:Zp-ffb}
\Gamma(Z'\to ff) =
\frac{g_{Zff}^2 M_{Z'} }{4 \pi }\sqrt{1-4 r_f} \left({g'}_{A}^2 (1-4 r_f)+{g'}_{V}^2 (2 r_f+1)\right)
\ee
The coefficients ${g'}^f_{V,A}$ are the couplings that appear in  eqs. (\ref{eq:gV-gA-SM-ext}) and 
$r_f = \frac{m_f^2}{M_{Z'}^2}.$ 

Finally, the decay width for the mode $Z' \to Zh$ is given by\cite{Abe:2011qe}:
\be\label{eq:Zp-Zh}
\Gamma(Z'\to Zh) &=&
\frac{g_{h Z Z'}^2 M_{Z'} r_{W Z}^{} }{64 \pi }
\sqrt{r_{h}^2-2 r_{h} (r_{Z}+1)+(r_{Z}-1)^2}\times
\nn\\
&&
\big(r_{h}^2-2 r_{h} (r_{Z}+1)+r_{Z} (r_{Z}+10)+1\big)
\ee
where $g_{h Z Z'}$ is shown in table (\ref{ta:ver-bos-higgs-fermio}), 
$r_{WZ} = \frac{M_W^2}{M_{Z}^2}, r_h = \frac{m_h^2}{M_{Z'}^2}$ and $r_Z =\frac{M_Z^2}{M_{Z'}^2}.$ 

In general, we shall use the following notation: $r_i = m_i^2/M_{Z'}^2$ and $r_{ij} = m_i^2/M_{j}^2.$ 

\subsection{$Z' \to f\bar{f} h$}

The amplitud for the decay $Z'\to f\bar{f}h$ procceds through the following Feynman diagrams,
\begin{figure}[H]
\vspace*{-2cm}
\includegraphics[width=6.5cm,angle =-90]{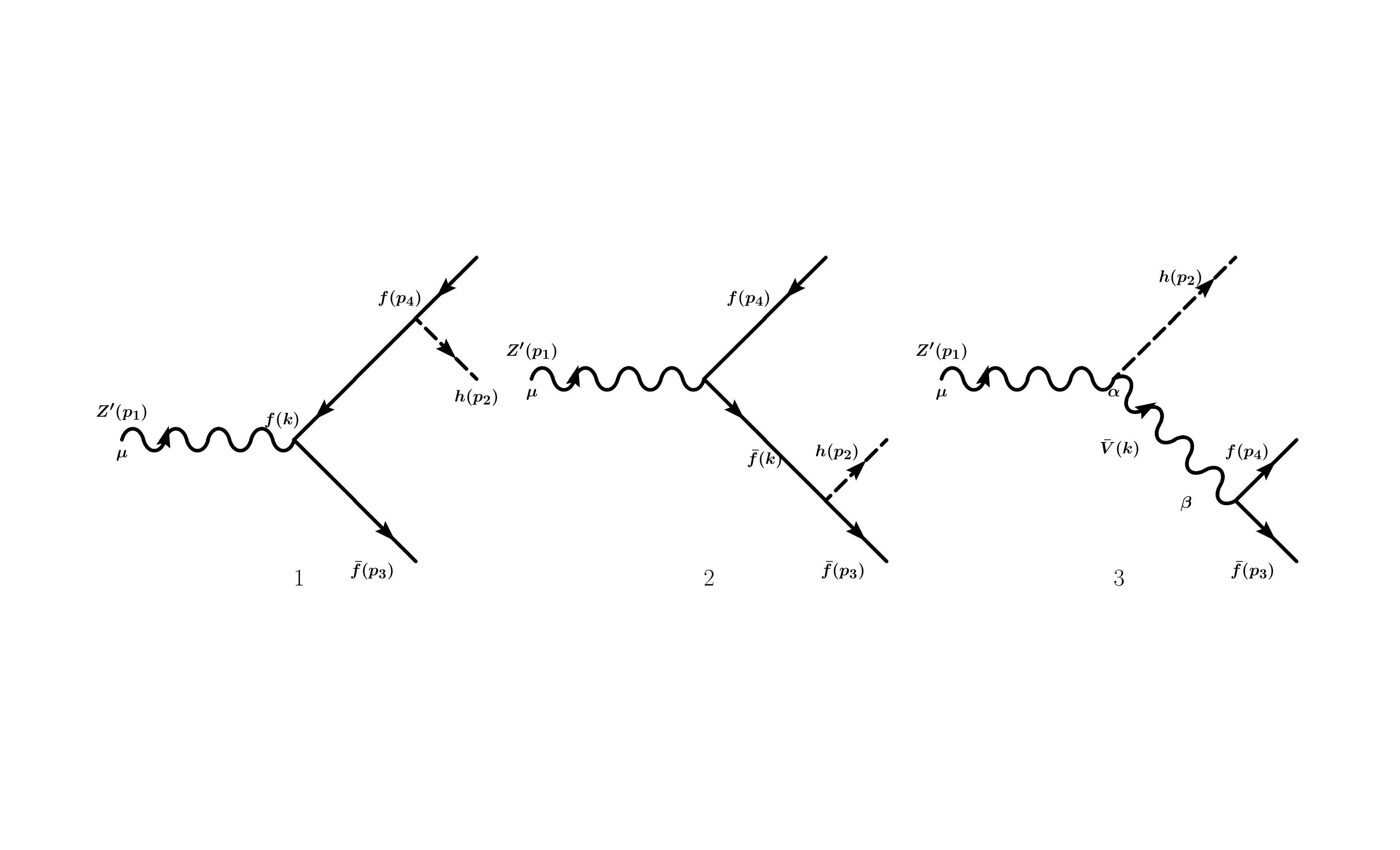}
\vspace*{-2cm}
\caption{Feynman diagrams for three bodies with involved femions in the final state.}
\end{figure}
The corresponding invariant amplitud is given by,
\be\label{eq:am-to2-reso}
\left|\overline{{\mathcal{M}}}_{123} \right|^2&=&= \sum_{i=1}^{3}\left|\overline{{\mathcal{M}}_i}\right|^2
	+2 \sum_{\stackrel{i,j = 1}{i\neq j}}^{3}{\Re}({{\mathcal{M}}}_i{{\mathcal{M}}}_j^*)\nn
\ee
By performing the calculations, we obtain the following result for the differential
decay width:

\be
\frac{d\Gamma(Z'\to f\bar{f}h)}{dxdy}&=&\frac{M_{Z'}}{256\pi^3}
\Bigg(g_{ffh}^2 g_{Z f f}^2 
f_1({g'}^f_A,{g'}^f_V;x, y)+\nn\\
&&
g_{f f h}^2 g_{Z f f}^2 
f_2({g'}^f_A,{g'}^f_V;x, y)+
 g_{Z f f}^2 g_{Z' Z h}^2
f_3(g^f_A,g^f_V;x, y)+\nn\\
&&
2\bigg(
g_{ffh}^2 g_{Zff}^2 
f_{12}({g'}^f_A,{g'}^f_V;x, y)+\nn\\
&&
g_{f f h}^{} g_{Z f f}^{} g_{Z f f}^{} g_{Z' Z h}^{}
f_{13}(g^f_A,{g'}^f_A,g^f_V,{g'}^f_V;x, y)+\nn\\
&&
g_{ffh}^{} g_{Zff}^{} g_{Z f f}^{} g_{Z' Z h}^{}
f_{23}(g^f_A,{g'}^f_A,g^f_V,{g'}^f_V;x, y)
\bigg)
 \Bigg)\label{eq:dbr-Zp-xy-ffh}
\ee
the integration limits for the scaled energy variables $(x,y) = (2E_h/M_{Z'},
2E_f/M_{Z'})$ are,
\be
2\sqrt{r_2} \leq x \leq 1 + r_2 - 4r_4
\hspace{10mm}
\mbox{and}
\hspace{10mm}
\frac{A - B}{2C}
\leq y \leq \frac{A + B}{2C}
\nn
\ee
where $A = (2- x_h) (1 + r_h- x), ~B =\sqrt{(x^2 - 4 r_h) (1 + r_h - x - 4r_f) (1 + r_h -x)}~$ and $C = (1 - x + r_h).$ The expressions for the functions $f_1,f_2,f_3, f_{12}, f_{13}, f_{23}$
are shown in Appendix \ref{ap:ap-m1m2m3}.

\subsection{$Z' \to W^+W^- h$}

The Feynman diagrams that contribute to the decay $Z' \to W^+W^-h$ are shown
next:
\vspace*{-2cm}
\begin{figure}[H]
\includegraphics[width=6.5cm,angle =-90]{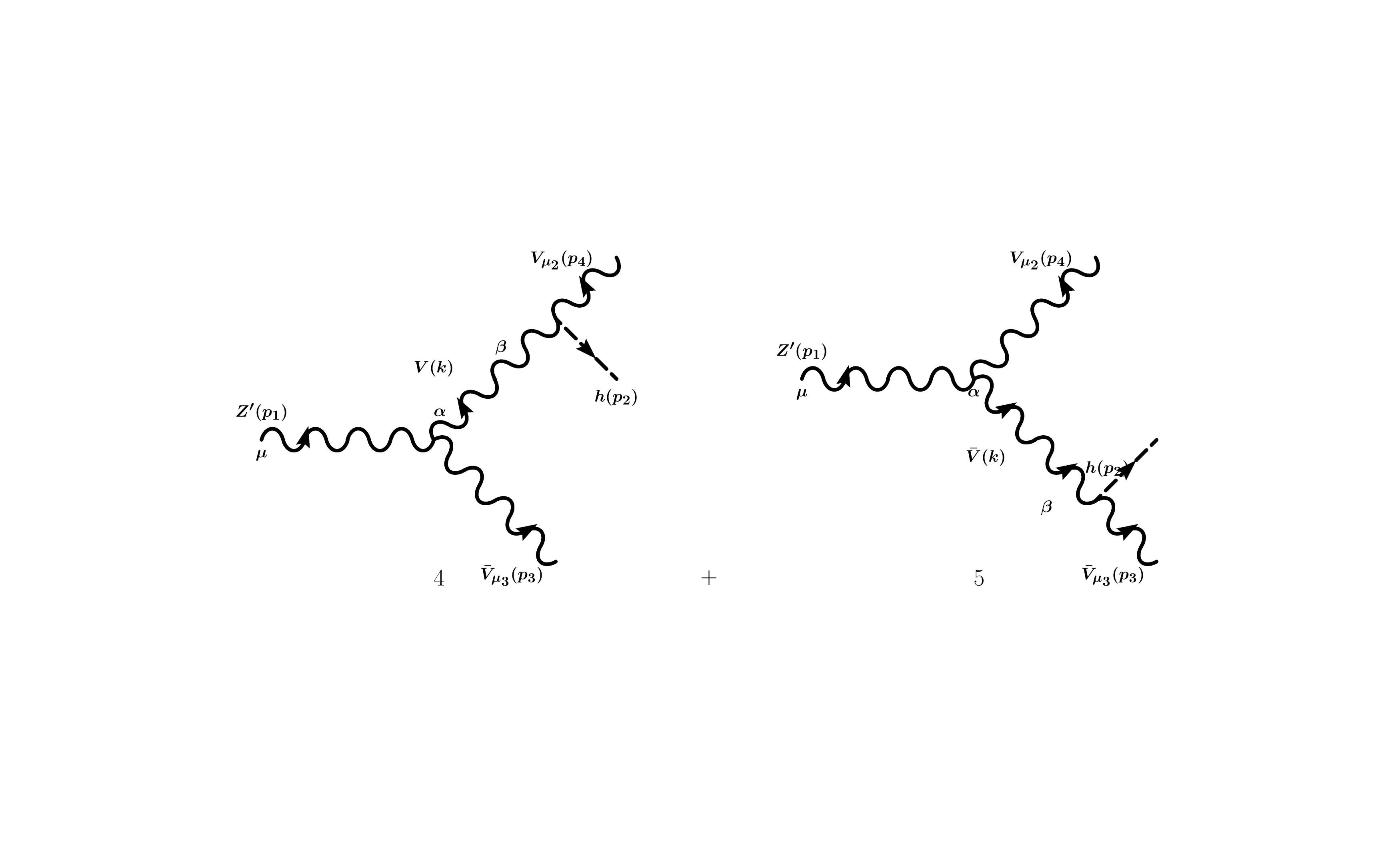}
\vspace*{-2cm}
\caption{Feynman diagrams for three bodies with involved gauge boson in the final state.}
\end{figure}
We have performed the calculation similarly to the last subsection, including the interference terms. The resulting differential decay width 
is written as: 
\be
\frac{d\Gamma(Z'\to W^+W^-h)}{dx dy}&=&\frac{M_{Z'} g_{Z'WW}^2 g^2 }{256\pi^3 r_W}
\Bigg(
f_4(x, y) + 
f_5(x, y)+
2f_{45}(x, y)
\Bigg)~~~\label{eq:dG-ZpWW}
\ee
for the definition of $f_{4,5,45}$ see the Appendix \ref{ap:ap-m4m5}.
The definitions for $(x, y)$ and their integration limits, are similar to the last calculation, 
with $x_f \to x_W.$ 
 
\subsection{Results for Br's}

The recent LHC results indicate that the Higgs boson mass is $m_h\simeq 125$ GeV 
\cite{CMS-bsn:2012gu,ATLAS-bsn:2012gk,erler-2012}, which is the nominal value we use in the following plots. 
We show the Branching ratio for different channels as a function of the $Z'$ gauge boson mass. 
We consider the $B-L$ scenarios for the fermion-$Z'$ couplings, with
$\theta' = 10^{-3}$ ($\theta' = 10^{-4}$).

We can see from figure \ref{fig:Br-GF-vp-1TeV} that the
  ${\Br}(Z'\to WW)$ is not dominant, because this process depends on
 the coupling $g_{Z'WW}$ (eq.\eqref{eq:Zp-WMWm}) which 
  decreases with $\theta'.$ For the ${\Br}(Z'\to f\bar{f})$ 
and ${\Br}{(Z'\to Zh)},$ the global factor coming from the
denominator in the second process is about ten times larger than the
one appearing in the first process, such that:
  ${\Br}(Z'\to f\bar{f})> {\Br}(Z'\to Zh).$ 
The decay $Z' \to t \bar{t} h$  depends directly on the couplings 
$g'_V$ and $g'_A$, in this case we use $g_1 \sim g'.$ The $Z' \to WW$
depends directly on $\theta'^2 \sim 1\times 10^{-6}$, so this is suppresed
too. The $Z' \to WWh$ process depends on the product $g_{Z'WW}
g_{WWh}$ and it inversely proportional too, in such way that as
$M_{Z'}$ grows then the corresponding width grows too.
\begin{figure}[H]
\vspace*{-1.cm}
\includegraphics[width=6.5cm,angle =-90]{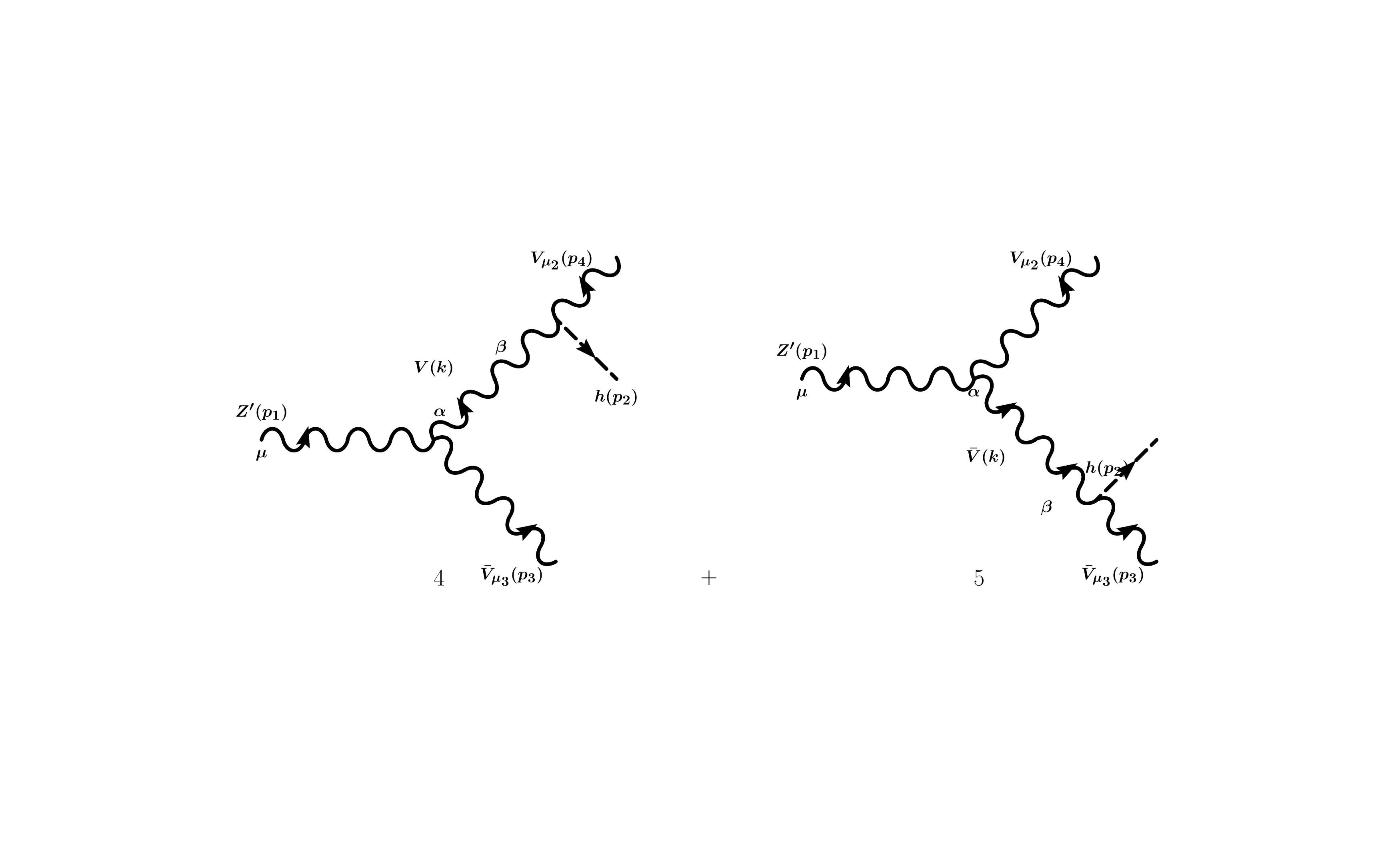}
\vspace*{-2cm}
\caption{Feynman diagrams for three bodies with involved gauge boson in the final state.}
\end{figure}
We have performed the calculation similarly to the last subsection, including the interference terms. The resulting differential decay width 
is written as: 
\be
\frac{d\Gamma(Z'\to W^+W^-h)}{dx dy}&=&\frac{M_{Z'} g_{Z'WW}^2 g^2 }{256\pi^3 r_W}
\Bigg(
f_4(x, y) + 
f_5(x, y)+
2f_{45}(x, y)
\Bigg)~~~\label{eq:dG-ZpWW}
\ee
for the definition of $f_{4,5,45}$ see the Appendix \ref{ap:ap-m4m5}.
The definitions for $(x, y)$ and their integration limits, are similar to the last calculation, 
with $x_f \to x_W.$ 
 
\subsection{Results for Br's}

The recent LHC results indicate that the Higgs boson mass is $m_h\simeq 125$ GeV 
\cite{CMS-bsn:2012gu,ATLAS-bsn:2012gk,erler-2012}, which is the nominal value we use in the following plots. 
We show the Branching ratio for different channels as a function of the $Z'$ gauge boson mass. 
We consider the $B-L$ scenarios for the fermion-$Z'$ couplings, with
$\theta' = 10^{-3}$ ($\theta' = 10^{-4}$).

\begin{figure}[H]
\vspace*{-1.cm}
\includegraphics[width=6.5cm,angle =-90]{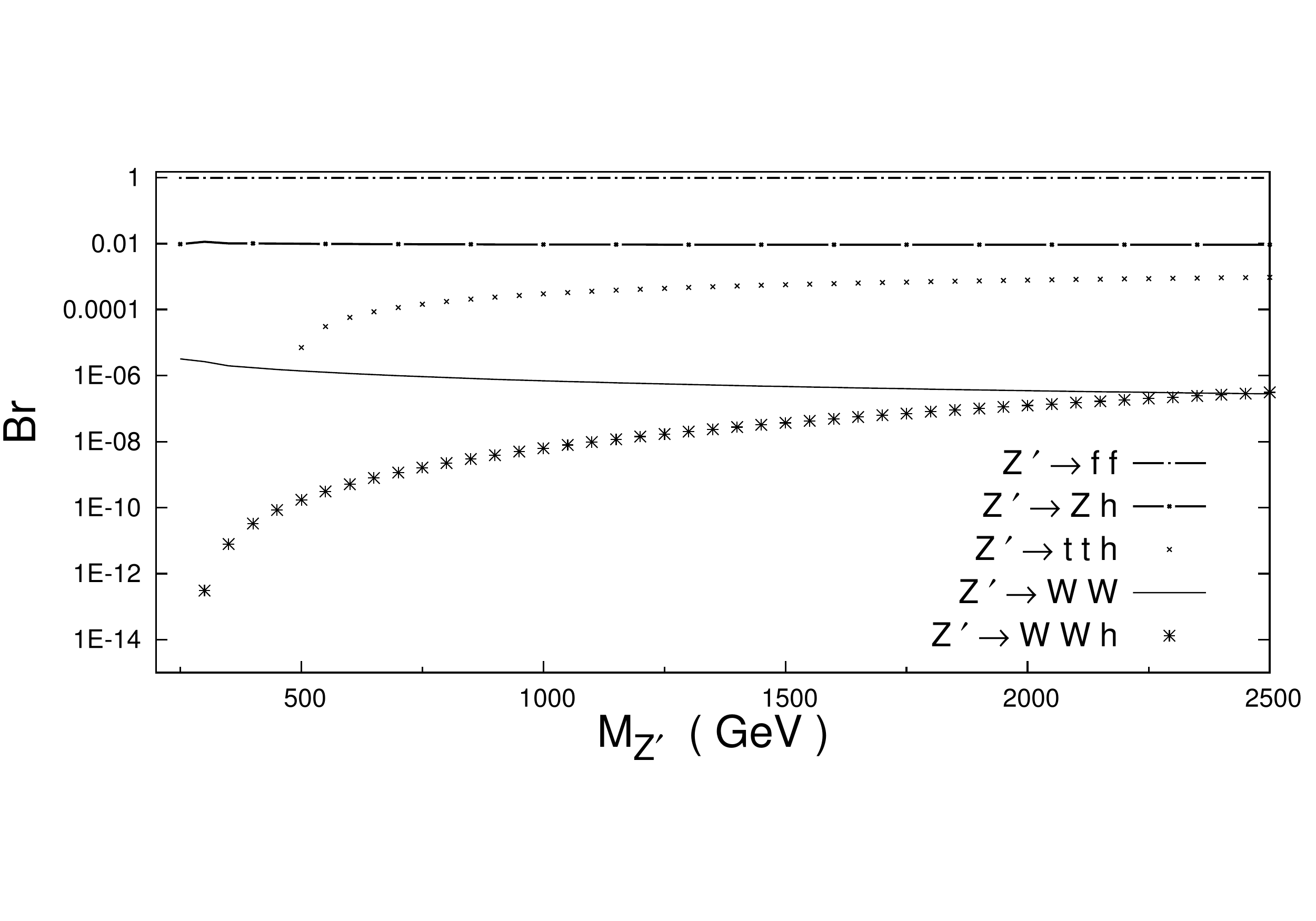}
\vspace*{-1.cm}
\caption{$v'=2 \TeV.$ \label{fig:Br-GF-vp-1TeV}}
\end{figure}

We can see from figure \ref{fig:Br-GF-vp-1TeV} that the
  ${\Br}(Z'\to WW)$ is not dominant, because this process depends on
 the coupling $g_{Z'WW}$ (eq.\eqref{eq:Zp-WMWm}) which 
  decreases with $\theta'.$ For the ${\Br}(Z'\to f\bar{f})$ 
and ${\Br}{(Z'\to Zh)},$ the global factor coming from the
denominator in the second process is about ten times larger than the
one appearing in the first process, such that:
  ${\Br}(Z'\to f\bar{f})> {\Br}(Z'\to Zh).$ 
The decay $Z' \to t \bar{t} h$  depends directly on the couplings 
$g'_V$ and $g'_A$, in this case we use $g_1 \sim g'.$ The $Z' \to WW$
depends directly on $\theta'^2 \sim 1\times 10^{-6}$, so this is suppresed
too. The $Z' \to WWh$ process depends on the product $g_{Z'WW}
g_{WWh}$ and it inversely proportional too, in such way that as
$M_{Z'}$ grows then the corresponding width grows too.
\begin{figure}[H]
\vspace*{-1.cm}
\includegraphics[width=6.5cm,angle =-90]{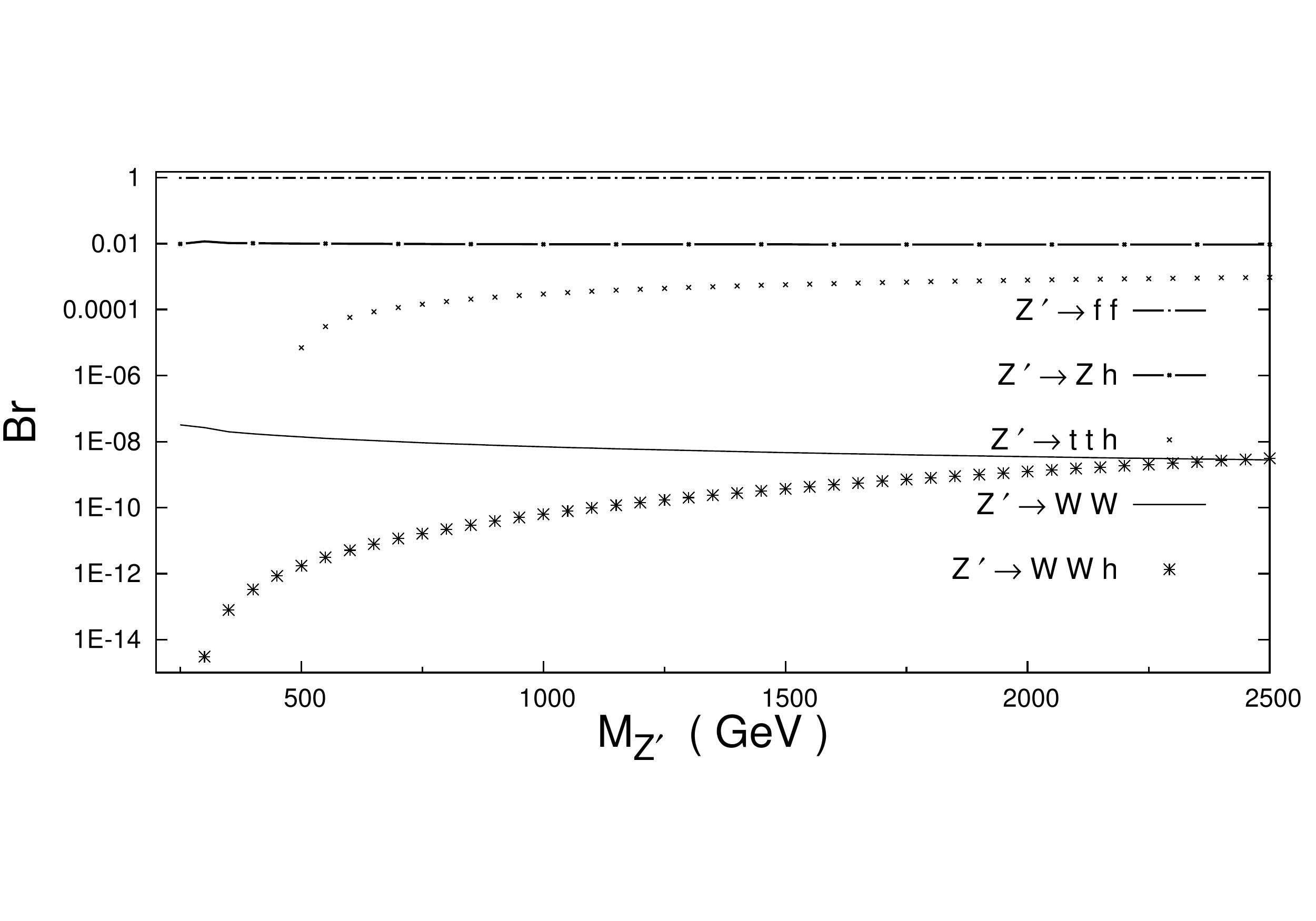}
\vspace*{-1.cm}
\caption{$v'=2 \TeV.$ \label{fig:Br-GF-vp-2TeV-tp-1-4}}
\end{figure}
In fig. \ref{fig:Br-GF-vp-1TeV}  we show the numerical results for $\theta' = 10^{-3}$, 
and for comparision we also show in fig. and \ref{fig:Br-GF-vp-2TeV-tp-1-4} the corrresponding ones for
 $\theta' = 10^{-4}$. We notice that there is a decrease in the decays $Z'\to WW, WWh$, whereas the 
decay $Z'\to ff, tth$ remmain essentially constant.

%
\section{The reaction $e^+ e^- \to Z', Z^*\to Z+h$ at ILC }
In this section we study the Higgs production at $e^+e^-$ colliders, through the 
reaction $e^+e^- \to Z, Z' \to Zh,$ taking into account the contribution of resonant ($Z'$) 
and non-resonant effects ($Z$). When $\sqrt{s}=M_{Z'}$ one expectes that the production
of $Z+h$ is dominated by the decay $Z'\to Zh$, however the vertex $ZZ'h$ is suppressed by the
mxiing angle $\theta'$, and therefore it is interesting to study the contribution from the
non-resonant amplitude.

\begin{figure}[H]
\vspace*{-1.cm}
\includegraphics[width=6.5cm,angle =-90]{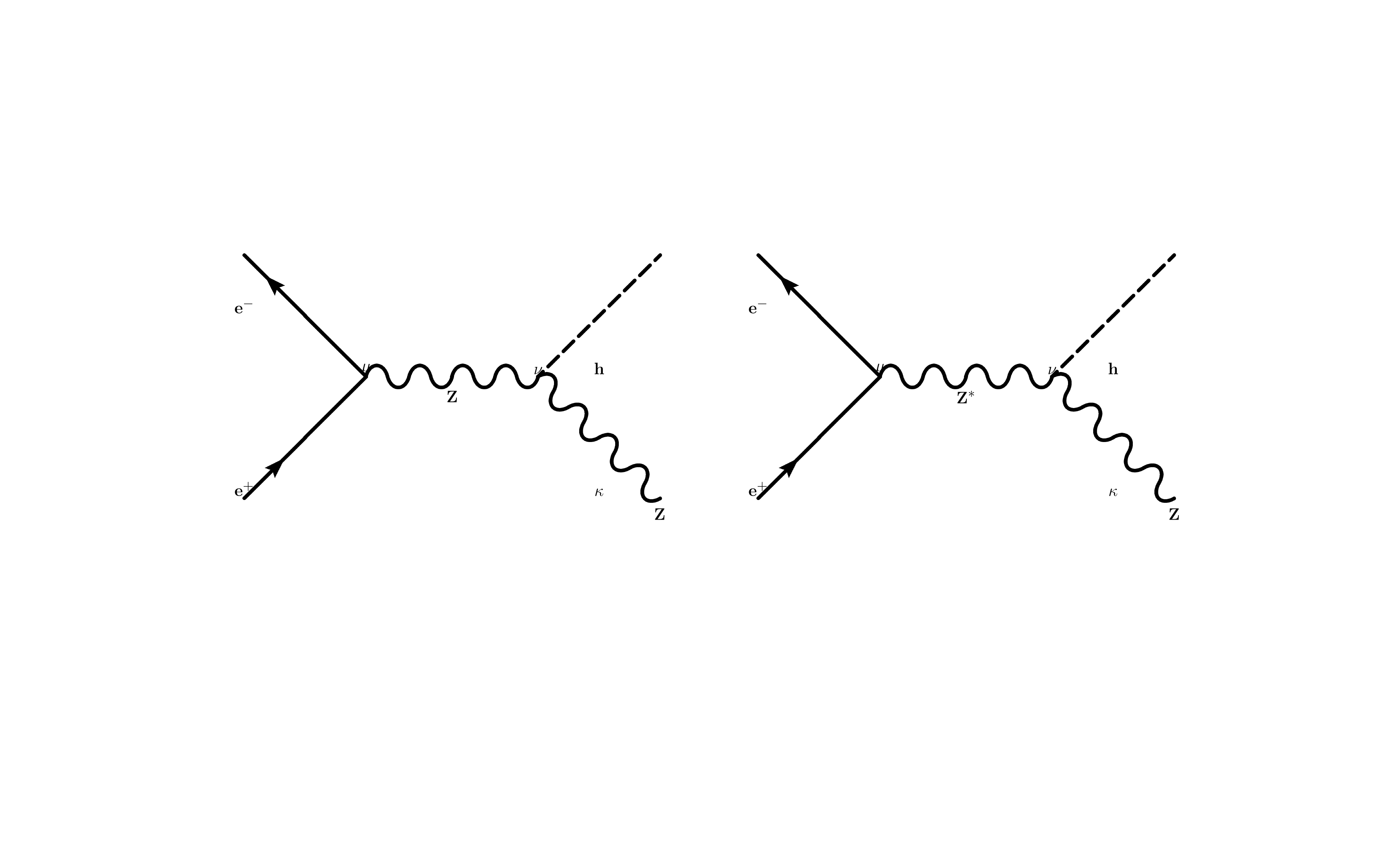}
\vspace*{-1.cm}
\caption{Feynman diagrams for $e^+ e^- \to Zh$ }\label{fd:fd-ee-Zp-Zh}
\end{figure}

The  Feynman diagrams for the reaction are shown in fig. \ref{fd:fd-ee-Zp-Zh},

the total amplitud is: ${\M} = {\M}_1 + {\M}_2.$ Where,
\be
\mathcal{M}_1&=&g_{ZZ'h} g_{Zff}g^{\nu\kappa}
\frac{ g_{\mu\nu}}{p_{Z'}^2 -  M_{Z'}^2 + iM_{Z'}\Gamma_{Z'}} \bar{u}
\gamma^\mu({g'}_V^f + {g'}_A^f\gamma^5) v{\epsilon^\kappa}^{*}\nn\\
\mathcal{M}_2&=&g_{ZZ'h} g_{Z'f\bar{f}}g^{\nu\kappa}
\frac{ g_{\mu\nu}}{p_{Z^*}^2 -  M_{Z}^2 + iM_{Z}\Gamma_{Z}} \bar{u}
\gamma^\mu(g_V^f + g_A^f\gamma^5) v{\epsilon^\kappa}^{*}.\nn
\ee
The resulting cross section is given by,

\be
\sigma (e^+e^- \to Z, Z' \to Zh) &=& \frac{ \bigg(\left(1-2 \sqrt{{r_h} {r_Z}}-{r_h}-{r_Z}\right) \left(1+2 \sqrt{{r_h}
   {r_Z}}-{r_h}+{r_Z}\right)\bigg)^{1/2} 
}{8\pi s r_{ZW}\sqrt{\frac{1}{4}\left(1 - 2{r_f} \right)^2-{r_f}^2}}\times\nn\\
&&\left[\frac{g_{Z'h}^2 g_{Z'f\bar{f}}^2 \bigg(
{{g'}_A^f}^2 R_{fZ}+{{g'}_V^f}^2 T_{fZ}
\bigg)}{2{c'}_r^2+ 2{r_{Z'}} {R_{Z'}}}
 +\frac{g_{Zh}^2 g_{Z f \bar{f}}^2 \bigg(
{{g}_A^f}^2 R_{fZ}+{{g}_V^f}^2 T_{fZ}
\bigg)}{2c_r^2 + 2r_Z R_{Z} } +\right.\nn\\
&&
\left.
\frac{{g_{Z'h}^{} g^{}_{Z' f \bar{f}} g_{Zh}^{} g^{}_{Z f \bar{f}}} \bigg(c'_r c_r + \sqrt{{r_{Z'}}{r_Z} {R_{Z'}}{R_{Z}}}\bigg)\bigg(
{g_A^f} {{g'}_A^f} R_{fZ}+{g_V^f} {{g'}_V^f} T_{fZ}
\bigg)}{\Big(c'_r c_r+ \sqrt{{r_{Z'}}{r_Z}{R_{Z'}}{R_{Z}}}\Big)^2 +\Big( \sqrt{{r_{Z'}}  {R_{Z'}}}c_r - c'_r \sqrt{{r_Z} {R_{Z}}}\Big)^2  }\right]~~~~~~~
\ee\label{eq:sigma-resonante}

where $R_{fZ} = (4-32 {r_f}) {r_Z}+1, ~~T_{fZ}=4 (4 {r_f}+1) {r_Z}+1, ~~c_r = 1-r_Z, ~~c'_r = 1-r_{Z'}$ and $g_{Zff},$ and $ g_{hZZ'}$ appear in eq. (\ref{eq:gV-gA-SM}) and in table (\ref{ta:ver-bos-higgs-fermio}). In this section, we have used $r_i = \frac{m_i^2}{s}, ~~r_{ij} = \frac{m_i^2}{m_{j}^2}, ~~R_Z = \frac{\Gamma_Z^2}{s}$ and $R_{Z'} = \frac{\Gamma_{Z'}^2}{s}.$ The $g_{V,A}'(g_{V,A})$ are the $Z'ff (Zff)$ couplings that appear in eq.\eqref{eq:L-Zff}.
\begin{figure}[H]
\includegraphics[width=6.5cm,angle =-90]{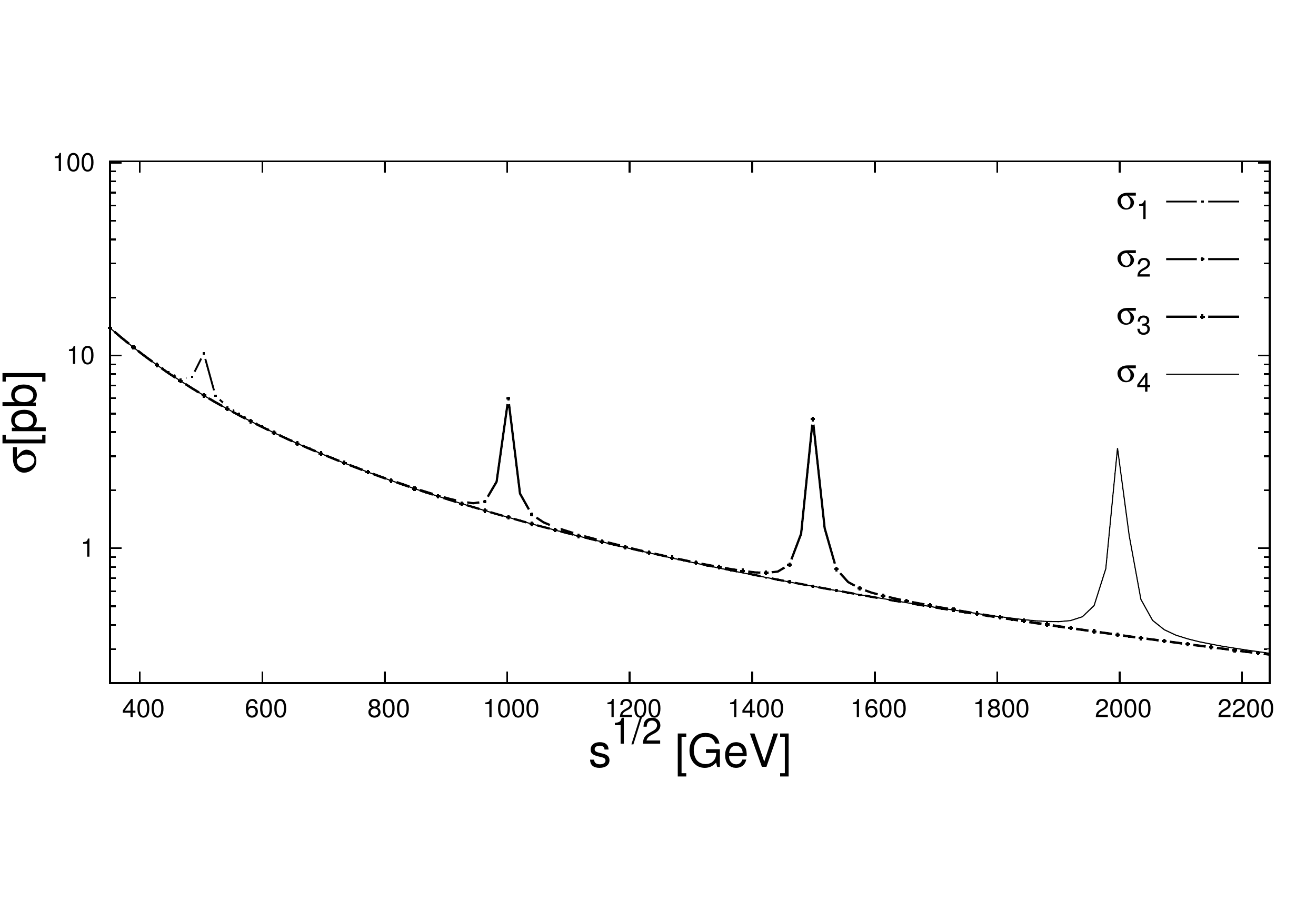}
\vspace*{8pt}
\caption{The cross sections for Higgs production with the $Z'$ in resonance.
 We use $v' = 2 \TeV,$ the intensity in resonance decreases when we raise the value for $v'$. $\sigma_{1-4}$ correspond to the cases of $M_{Z'} = 0.5,1,1.5,2 \TeV$,respectively. \label{fig:proces-resonan}}
\end{figure}

We can see from the fig. \ref{fig:proces-resonan}, that when $\sqrt{s} = M_{Z'},$ the resonant effect dominates the Higgs production, despite the fact that the coupling $ZZ'h$ is suppresed.
\section{Conclusions}
Models containing an extra gauge boson often include  
a rich spectrum of Higgs bosons, with properties 
that deviate from the Standard Model (SM) case. 
Such Higgs bosons could be searched using standard mechanisms, similar to the SM.
However, the existence of such new gauge bosons could also be used to 
provide new production mechanisms, which could probe the non-standard origin
of the Higgs couplings.

In this paper we have studied how the $Z'$ arising from a model with a extra 
$U(1)'$ could be used as a source of Higgs bosons.  For such a task,
we concentrate in the $Z' \to \bar t t h$ channel, and we found a
{\Br} for this mode $10^{-2}$ far a wide range  of the parameter space, which could be
within the reach for both accelerators LHC and ILC.
To appreciate the importance of the resonant effect for Higgs
production one can see the fig.\ref{fig:proces-resonan} when $\gamma$
is significant.

Further, at the next linear collider (ILC) one could achieve luminosites $\sim
{{\mathcal{O}}}({\ab}^{-1})$ at $\sqrt{s} \sim {{\mathcal{O}}}(\TeV)$
\cite{Simon:2012ik, Jiang:2012bf}, which could give a number of events:
 $N=\sigma({e^+e^-\to Z, Z' \to Zh})_{\tiny{M_Z' = 1 \TeV}} \times
{\mathcal{L}}_{NLC} \sim 1 {\pb} \times 1 {\ab^{-1}} \sim 10^{6},$
which seems feasible to be detected and it will be possible to perform
precision measurements for $Z'$ and Higgs.
In models with an extra Higgs doublet, it could be possible to get larger coulings of the
Higgs boson with b-quarks, and this will also allow  to study the decay $Z' \to bbh$,
 which could be detactable even at LHC \cite{Balazs:1998nt}.

\section*{Acknowledgements}

We acknowledge support from CONACYT-SNI (Mexico). We thank Enrique D\'iaz for reading the 
manuscript and his comments. 

\appendix
\section{Appendices}\label{ap:ap-m1m2m3}
The next equations are used to obtain the $\Gamma(Z' \to f\bar{f}h)$ in eq.\eqref{eq:dbr-Zp-xy-ffh}, 
{\scriptsize{
\be
f_1({g'}^f_A,{g'}^f_V;x, y)
&=&-\frac{2}{(x+y-1)^2}
\Bigg({g'}_{A}^2 \bigg(32 
{r}_{f}^2+4 {r}_{f} \left(-2 {r}_{h}+x^2+2 x (y+1)+y 
(y+2)-5\right)+\nn\\
&&
(x+y) (-{r}_{h} (x+y-2)+x (x+y-3)+y)+{r}_{h}+x-2 
y+1\bigg)+\nn\\
&&
{g'}_{V}^2 \bigg(-16 {r}_{f}^2+4 {r}_{f} 
\left({r}_{h}+x^2+2 x (y-2)+(y-4) y+1\right)+\nn\\
&&
(x+y) (-{r}_{h} (x+y-2)+x 
(x+y-3)+y)+{r}_{h}+x-2 y+1\bigg)\Bigg)
\nn\\
f_2({g'}^f_A,{g'}^f_V;x, y)
&=&\frac{2}{(y-1)^2}
\Bigg({g'}_{V}^2 \bigg(r_{h} (8 r_{f}+(y-2) y-1)-(y-1) 
(x y+x+y-1)\bigg)\nn\\
&&
-{g'}_{A}^2 \bigg(r_{h} (4 
r_{f}-(y-2) y+1)+(y-1) (x y+x+y-1)\bigg)\Bigg)
\nn
\ee
}}
{\scriptsize{
\be
f_3(g_A,g_V;x, y)
&=&\frac{2 r_{W} }{(-r_{h}+r_{Z}+x-1)^2}
\Bigg(g_{V}^2 \bigg(4 
r_{f}+r_{h}-x y-x-y^2+2 y+1\bigg)-\nn\\
&&
g_{A}^2 \bigg(8 
r_{f}-r_{h}+(x-2) 
y+x+y^2-1\bigg)\Bigg)\nn
\ee
}}
and the interference terms,
{\scriptsize{
\be
f_{12}({g'}^f_A,{g'}^f_V;x, y)
& = & \frac{2 }{(y-1) (x+y-1)}
\Bigg({g'}_{A}^2 \bigg(2 {r}_{f} (4 
{r}_{h}+(x+4) y+x-4)-2 {r}_{h}^2-\nn\\
&&
{r}_{h} \left(x (y-3)+(y-1)^2\right)+(x+1) (y-1) (x+y-1)\bigg)+\nn\\
&&
{g'}_{V}^2 \bigg(2 {r}_{f} \left(-4 {r}_{h}+x^2+(x-2) y+x+2\right)+2 {r}_{h}^2+\nn\\
&&
{r}_{h} 
\left(x (y-3)+(y-1)^2\right)-(x+1) (y-1) (x+y-1)\bigg)\Bigg)
\nn\\
f_{13}(g_A,{g'}^f_A,g_V,{g'}^f_V;x, y)
& = & -\frac{2  \sqrt{{r}_{f}}\sqrt{r_{W}} 
}{(x+y-1) (-{r}_{h}+r_{Z}+x-1)}
\Bigg({g}_{A} {g'}_{A} \bigg(16 {r}_{f} 
+
2 
y^2-4 {r}_{h}+3 x y+\nn\\
&&
x (x+3)-6\bigg)+
{g}_{V} {g'}_{V}\bigg(2 {r}_{h}+(x+y-3) (x+2 y)-8 
{r}_{f}\bigg)\Bigg)
\nn\\
f_{23}(g_A,{g'}^f_A,g_V,{g'}^f_V;x, y)
& = & \frac{2  \sqrt{{r}_{f}} \
\sqrt{{r}_{W}} }{(x+y-1) ({r}_{h}-{r}_{Z}-x+1)}
\Bigg({g}_{A} {g'}_{A} \bigg(2 {r}_{h}+(x+4) y+x-4\bigg)+\nn\\
&&
{g}_{V} 
{g'}_{V} \bigg(-4 {r}_{h}+(x-2) y+x+2\bigg)\Bigg)
\nn
\ee
}}

 \section{Appendices}\label{ap:ap-m4m5}
The next equations are used to obtain the $\Gamma(Z' \to W^+W^-h)$ in eq.\eqref{eq:dG-ZpWW},
{\scriptsize{
\be
f_4(x, y)
& = & \frac{1}{4 (x+y-1)^2}
\Bigg(4 (x-y-14) r_{W}^3+4 \left(3 y^2+6 x 
y-20 y+2 (x-5) x+17\right) r_{W}^2+
\bigg(x^3+\nn\\
&&(3 y-2) x^2+(y-1) (7 y-3) 
x+(y-1)^2 (5 y-14)\bigg) r_{W}+(x-y+1) (y-1)^2-
r_{h}^2 
\bigg(y-x^2+\nn\\
&&x+r_{W} (x-y+2)-1\bigg)+2 r_{h} \left(8 
r_{W}^2-\left(y^2+3 x y-7 y+x (2 x-7)+6\right) 
r_{W}-(y-1)^2\right)\Bigg)
\nn\\
f_5(x, y)
& = & \frac{1}{16 (y-1)^2}
\Bigg(8 r_{h}^3+\left(x^2+4 (2 y-3) x+16 
(y-1)^2+r_{W} (-8 x+8 y-52)\right) r_{h}^2+\nn\\
&&
2 \left(8 (x-y+7) 
r_{W}^2-4 (y-1) (3 x+6 y-7) r_{W}-\left(x^2+4 (y-1) x+4 (y-1)\right) 
(y-1)\right) r_{h}+\nn\\
&&
x (x+4) (y-1)^2+16 r_{W}^3 (x-y-14)-8 r_{W}^2 
\left(x^2-4 x-8 y^2+14 y-6\right)+\nn\\
&&4 r_{W} \left((2 y-1) x^2+
4 (y-1) y 
x+(y-1)^2 (2 y-1)\right)\Bigg)\nn
\ee
}}
and 
{\scriptsize{
\be
f_{45}(x, y)
& = & -\frac{1 }{8 (-2 r_{h}+2 r_{W}+y-1) (x+y-1)}
\Bigg(r_{h}^3+\bigg(x^2+4 y x+
2 
r_{W} (x-y-8)  +\nn\\
&&3(1-2x-y) \bigg) r_{h}^2+\bigg(4(y-x+22) 
r_{W}^2-
\left(5 x^2+16 y x-24 x+12 y^2-40 y+28\right) r_{W}-\nn\\
&&(y-1) 
\left(x^2+5 y-5\right)\bigg) r_{h}+
(2 x-y+1) (y-1)^2+8 r_{W}^3 
(x-y-14)+4 r_{W}^2 \bigg(7 y^2-27 y+\nn\\
&&x (6 y-8)+20\bigg)+r_{W} 
\left(x^3+(5 y-3) x^2+\left(6 y^2-4 y-2\right) x+6 (y-2)\right)\Bigg) \nn
\ee
}}

\bibliographystyle{plain}
\bibliography{biblio-zprime-2}

\end{document}